\documentclass[oneside,twocolumn]{article}

\usepackage{lmodern}
\usepackage[T1]{fontenc} 
\usepackage{microtype} 
\usepackage[english]{babel} 
\usepackage{graphicx}

\usepackage{epsfig}

\usepackage[hmarginratio=1:1,top=18mm,bottom=18mm,left=18mm,right=14mm,columnsep=20pt]{geometry}
\usepackage[small,labelfont=bf,up,up]{caption} 
\usepackage{booktabs} 
\usepackage{lettrine} 

\usepackage{enumitem} 
\setlist[itemize]{noitemsep} 
\usepackage{tabularx,ragged2e}

\usepackage{abstract} 

\usepackage{titlesec} 
\titleformat{\section}[block]{\large\scshape\centering}{\thesection.}{1em}{} 
\titleformat{\subsection}[block]{\large\scshape\centering}{\thesubsection.}{1em}{}

\usepackage{fancyhdr} 
\usepackage{titling} 
\usepackage{hyperref} 

\pretitle{\begin{center}\LARGE\bfseries} 
	\posttitle{\end{center}} 
\title{Large-Scale Fluctuations in the Number Density of Galaxies
	in Independent Surveys of Deep Fields}
\author{\textsc{\bfseries Shirokov S. I.$^{1}$}\thanks{E-mail: arhath.sis@yandex.ru}, \textsc{\bfseries Lovyagin N. Yu.$^{1}$}, \textsc{\bfseries Baryshev Yu. V.$^{1}$}, and	\textsc{\bfseries Gorokhov V. L.$^{2}$}\\
	\normalsize \itshape $^{1}$Saint-Petersburg State University, University Embankment 7-9, St. Petersburg, 199034 Russia\\
	\normalsize \itshape $^{2}$Saint-Petersburg State University of Architecture and Civil Engineering, St. Petersburg, Russia\\
	\\
	\small	ISSN 1063-7729, Astronomy Reports, 2016, Vol. 60, No. 6, pp. 563-578.
	\small	PleiadesPublishing, Ltd., 2016.\\
	\small	Original Russian Text S.I. Shirokov, N.Yu. Lovyagin, Yu.V. Baryshev, V.L.	Gorokhov, 2016,\\
	\small published in Astronomicheskii Zhurnal, 2016, Vol. 93, No. 6, pp. 546-561.\\
	\scshape {\url{http://link.springer.com/article/10.1134/S1063772916040107}}
}
\date{Received June 7, 2015; in final form, October 20, 2015} 
\renewcommand{\maketitlehookd}{%
	\begin{abstract}
		\noindent New arguments supporting the reality of large-scale fluctuations in the density of the visible matter in deep galaxy surveys are presented. A statistical analysis of the radial distributions of galaxies in the COSMOS and HDF-N deep fields is presented. Independent spectral and photometric surveys exist for each field, carried out in different wavelength ranges and using different observing methods. Catalogs of photometric redshifts in the optical (COSMOS-Zphot) and infrared (UltraVISTA) were used for the COSMOS field in the redshift interval $0.1 < z < 3.5$, as well as the zCOSMOS (10kZ) spectroscopic survey and the XMM-COSMOS and ALHAMBRA-F4 photometric redshift surveys. The HDFN-Zphot and ALHAMBRA-F5 catalogs of photometric redshifts were used for the HDF-N field. The Pearson correlation coefficient for the fluctuations in the numbers of galaxies obtained for independent surveys of the same deep field reaches $R = 0.70 \pm 0.16$. The presence of this positive correlation supports the reality of fluctuations in the density of visible matter with sizes of up to 1\,000 Mpc and amplitudes of up to 20\% at redshifts $z \sim 2$. The absence of correlations between the fluctuations in different fields (the correlation coefficient between COSMOS and HDF-N is $R = -0.20 \pm 0.31$) testifies to the independence of structures visible in different directions on the celestial sphere. This also indicates an absence of any influence from universal systematic errors (such as "spectral voids"), which could imitate the detection of correlated structures.
		
		{\bf Key words:} Cosmology: observations; large-scale structure of Universe
	\end{abstract}
}

\begin{document}
	
	\maketitle
	
	\section{Introduction}

Modern observational cosmology has led to the discovery of very large structures with scales of order 100 Mpc in the spatial distribution of galaxies in the local Universe, at redshifts $z \sim 0.1$, and also in the spatial distribution of quasars at redshifts $z \sim 2$ $[1-6]$. Over the last decade, observations of the largescale structure of the Universe [7] have moved from groups and clusters of galaxies with sizes of the order of 1 Mpc to structures with sizes of $\sim 100$ Mpc (SDSS superclusters, in particular, the Sloan Great Wall, with a size of 420 Mpc [1]). The mass distribution can be determined independently from analyses of the proper motions of galaxies. Tully et al. [2] recently discovered a coherent motion of galaxies forming the Laniakea (Local) Supercluster with a diameter of 160 Mpc. Groups of quasars with scales of $10-100$ Mpc have also been found, beginning with the study of Komberg et al. [3, $4-6$]. 

Modern multi-band photometric deep galaxy surveys can be used to study the spatial distribution of galaxies at redshifts $0.3-3$, which has led to the detection of inhomogeneities on scales up to 1\,000 Mpc $[8-10]$. A comparison of the wideangle Sloan Digital Sky Survey (SDSS) and the COSMOS pencil-beam survey is shown in Fig. 1, together with the radial distribution of the number of galaxies.

Fluctuations in the numbers of galaxies in neighboring volume elements of a pencil-beam survey are due to the presence of Poisson noise (the discreteness of the sample), systematic observational errors (selection effects), and the presence of large-scale structure (the "cosmic variance"), which plays an important role in comparisons of models with the observations. The main difficulty in distinguishing real density fluctuations is the possibility of hidden selection effects that are present in each galaxy survey, which can imitate large-scale inhomogeneity of the galaxy distribution.

\begin{figure*}
	\centering
	\includegraphics[scale=0.35,clip]{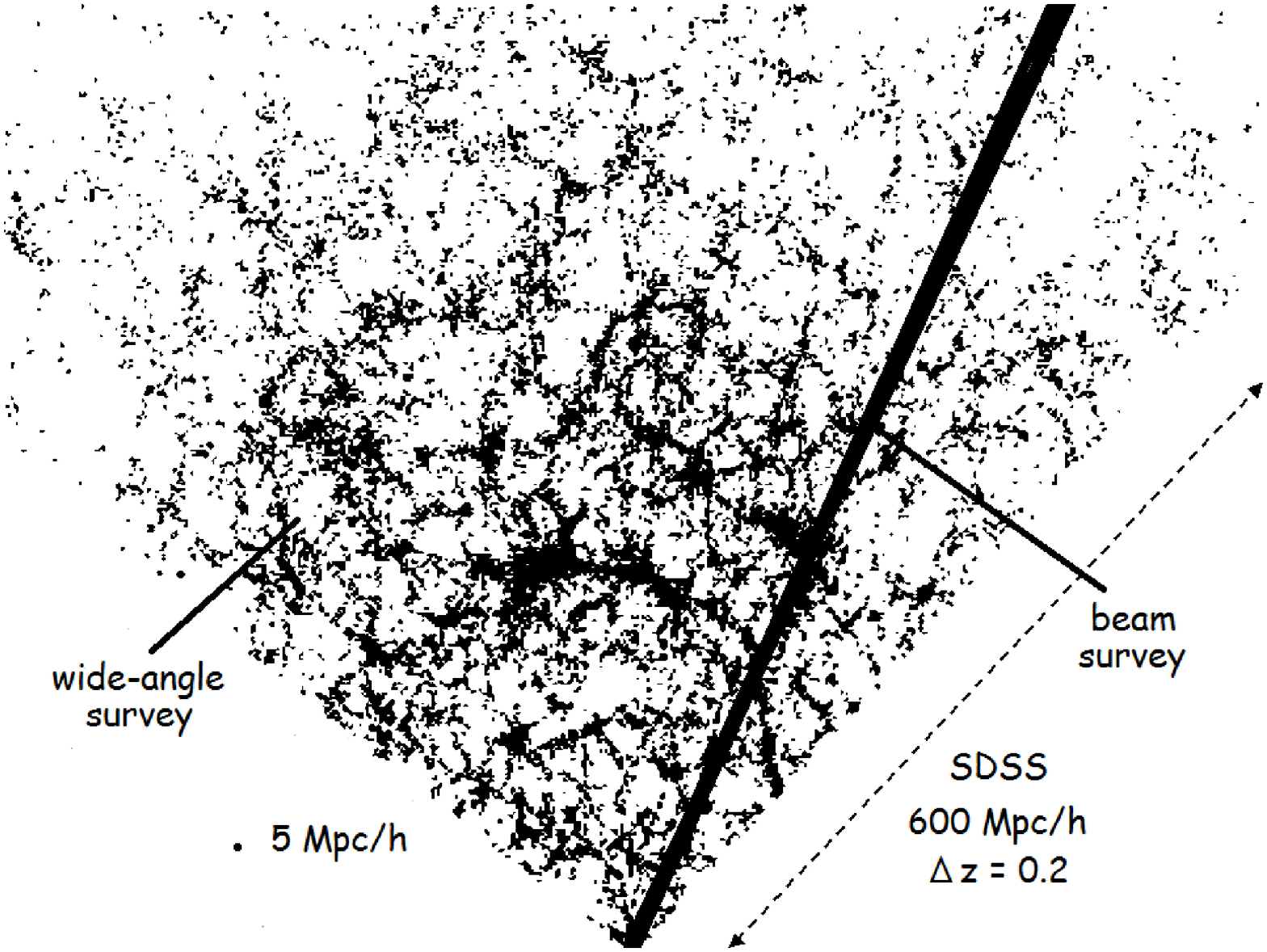}
	\hfill
	\includegraphics[scale=0.35,clip]{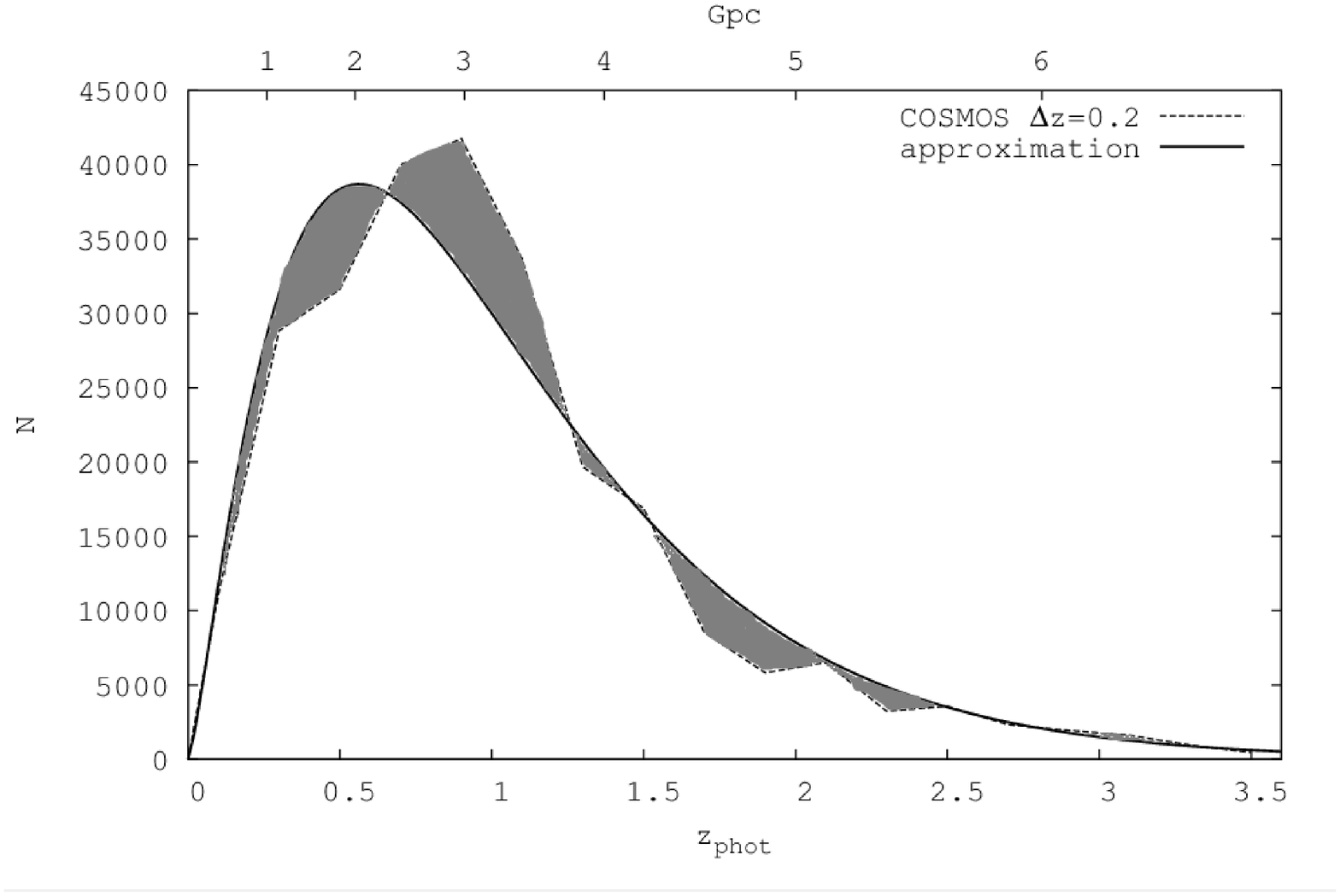}
	\label{SDSS}
	
	\caption{ Upper: distribution of galaxies in the SDSS. The one-square-degree COSMOS deep field is marked by the dark strip. Lower: fit of the radial distribution of the number of galaxies using a uniform distribution. The gray area highlights regions where there are deficits or excesses of galaxies.}
\end{figure*}

In the current study, we present new arguments supporting the reality of large-scale fluctuations in the matter density in deep galaxy surveys.

\section{Method}

\subsection{Estimation of the Amplitude	and Scale of Fluctuations}
We adopted the method proposed in [8, 9] and developed in [10] as the basis for our analysis. When
dealing with large redshifts ($z \sim 2$), it is necessary to use the exact formulas for a standard Friedmann model when calculating the metric distances (r(z)) and ages (t(z)) of galaxies ([7, Chapter 7]):	

\begin{equation}
 \label{R_metric}
 r(z)=\frac{c}{H_0}\int_{0}^z\frac{dy}{h(y)},
\end{equation}
where $c$ is the speed of light, $H_0=72$ km/s/Mpc, $\Omega_m^0=0.3$ and $\Omega_v^0=0.7$, and 
\begin{equation}
 \label{Age}
 t(z)=\frac{1}{H_0}\int_z^\infty \frac{dy}{(1+y)h(y)},
\end{equation}
where
\begin{equation}
 \label{h(y)}
 h(y)=\sqrt{\Omega_v^0+\Omega_m^0(1+y)^3-(1-\Omega_{tot}^0)(1+y)^2}.
\end{equation}

According to [11, 12], the distribution of galaxies in a deep survey can be approximated by the empirical expression
\begin{equation}
 \label{N_mod}
 N_{mod}(z,\Delta z)=Az^\alpha e^{\left( -z/z_0\right)^\beta}\Delta z,
\end{equation}

where $ N_{mod}(z,\Delta z)$ is the number of galaxies in the interval $(z, z+\Delta z)$. The parameters $\alpha$, $\beta$ and $z_0$ are determined from a least-squares fit and A is a constant that normalizes the integral of the model to the total number of galaxies $\int$N$_{mod}$dz=N$_{tot}$. Formula (4) was also confirmed using model galaxy samples in [13]. Deep galaxy surveys characteristically have small angular sizes. As was shown in $[14-16]$, the expected theoretical dispersion in the relative density fluctuations $\sigma^2$ is the sum of the dispersions of the correlation structures $\sigma^2_{corr}$ and the Poisson noise $\sigma_p^2$,

\begin{equation}
 \label{Sigma}
 \sigma^2(z,\Delta z)=\sigma_{corr}^2+\sigma_p^2 \;.
\end{equation}
Poisson noise has a dispersion of
\begin{equation}
 \label{Sigma_p}
 \sigma^2_P=\frac {\left\langle N^2\right\rangle-\left\langle N\right\rangle^2}{\left\langle N\right\rangle^2}=\frac{1}{\left\langle N\right\rangle},
\end{equation}
When the number of points in the considered volume is sufficiently large, the fluctuations of a uniformdistribution become insignificant ($\sigma_p^2 \sim N^{-1/2}$). Therefore, the total dispersion is dominated by the socalled cosmic variance $\sigma^2_{corr}$, which is expressed in [14, 17] in terms of the spatial correlation function $\xi(r)$ using the formula
\begin{equation}
 \label{Sigma_Int}
 \sigma_{corr}^2(V)=\frac{1}{(1+z)V^2}\int_V dV_1\int_V dV_2\xi(|r_1-r_2|)
\end{equation}
where $V = V (z, \Delta z)$ is the volume of the integrated region, corresponding to the interval $(z, z+\Delta z)$, the factor $1/(1 + z)$ takes into account the linear growth in the fluctuation amplitude with time, and, in the case of a power-law density distribution, the correlation function has the form
\begin{equation}
 \label{xi}
\xi(r)=\left( \frac{r_0}{r}\right)^\gamma
\end{equation}
\begin{equation}
 \label{Sigma_est}
 \sigma_{corr}^2(z,\Delta z)=\frac{J_2}{1+z}\left( \frac{r_0}{r}\right)^\gamma
\end{equation}
The amplitude of the deviations of the observed galaxy distribution $ N_{obs}(z,\Delta z)$ from a uniform distribution $ N_{mod}(z,\Delta z)$ is specified by the ratio
\begin{equation}
 \label{delta_obs}
 \delta_{obs}(z, \Delta z)=\frac{\Delta N_{obs}}{N_{mod}} =
 \frac {N_{obs}-\left\langle N_{mod}\right\rangle}{\left\langle N_{mod}\right\rangle},
\end{equation}
\begin{table*}[h]
	\centering
	\caption{Reduced disperson of the COSMOS field.}
	\begin{tabularx}{\textwidth}{XXXXXXXXX}
		\hline \hline	
		$\Delta z$   &  COSMOS	&  UVISTA  &  zCOSMOS	&  ALH-F4	&  XMM$_{phot}$\\
		\hline
		0.3  &  10.35	&  1.14    &            &  4.03	    &  1.31\\[3pt]
		0.2  &  6.93	&  4.09	   &  5.78	    &  4.6	    &  5.21\\[3pt]
		0.1  &  8.68	&  3.79	   &  5.00	    &  3.96	    &  5.26\\
		0.05 &  7.26	&  3.09	   &  5.16\\[3pt]
		\hline
	\end{tabularx}
	
	\label{rv_tab}
\end{table*}

\begin{table*}[h!]
	\caption{Least-squares parameters ($\alpha, \beta, z_0$) and the sum of the squared deviations ($\Sigma$) for all samples and all bins }
	\begin{tabular}{llllllllllllllll}
		\hline 
		$\Delta z$	&	$\alpha$	&	$\beta$	&	$z_0$	&	$\delta$	&	$\Sigma$	&	$\alpha$	&	$\beta$	&	$z_0$	&	$\delta$	&	$\Sigma$	&	$\alpha$	&	$\beta$	&	$z_0$	&	$\delta$	&	$\Sigma$	\\
		\hline
		&	\multicolumn{5}{c}{COSMOS $z_{max} = 3.6, N\approx2\cdot 10^5$}		&	\multicolumn{5}{c}{C\&U $z_{max} = 3.6, N\approx5\cdot 10^5$}		&	\multicolumn{5}{c}{ALH-F5 $z_{max} = 2.4, N\approx1\cdot 10^4$} 	\\
		
		0.05	&	0.60	&	1.51	&	1.13	&	0.1\%	&	14.21	&	0.49	&	1.49	&	1.13	&	0.2\%	&	5.87	&	1.15	&	1.96	&	0.94	&	0.2\%	&	4.72	\\
		0.1	&	0.84	&	1.27	&	0.83	&	0.2\%	&	4.66	&	0.68	&	1.32	&	0.91	&	0.5\%	&	2.01	&	1.26	&	1.87	&	0.88	&	0.6\%	&	1.71	\\
		0.2	&	1.36	&	0.94	&	0.38	&	1.4\%	&	0.70	&	1.06	&	1.06	&	0.55	&	2.8\%	&	0.36	&	1.19	&	1.93	&	0.92	&	1.4\%	&	0.69	\\
		0.3	&	1.37	&	0.94	&	0.38	&	1.8\%	&	0.57	&	1.22	&	1.00	&	0.46	&	4.2\%	&	0.24	&	0.95	&	2.33	&	1.10	&	5.1\%	&	0.19	\\
		&	\multicolumn{5}{c}{UVISTA $z_{max} = 3.6, N\approx2\cdot 10^5$}		&	\multicolumn{5}{c}{XMM$_{phot}$ $z_{max} = 3.6, N\approx2\cdot 10^3$} 	&	\multicolumn{5}{c}{HDF-N $z_{max} = 3.6, N\approx2\cdot 10^3$} 		\\
		0.05	&	0.32	&	1.66	&	1.33	&	0.4\%	&	2.71	&	0.40	&	2.23	&	2.00	&	0.1\%	&	8.73	&	1.32	&	0.44	&	0.05	&	0.1\%	&	14.69	\\
		0.1	&	0.46	&	1.54	&	1.18	&	1.1\%	&	0.95	&	0.52	&	2.02	&	1.80	&	0.5\%	&	1.96	&	1.32	&	0.44	&	0.05	&	0.2\%	&	6.03	\\
		0.2	&	0.58	&	1.43	&	1.04	&	3.7\%	&	0.27	&	0.56	&	1.99	&	1.75	&	2.2\%	&	0.45	&	1.33	&	0.54	&	0.14	&	0.7\%	&	1.51	\\
		0.3	&	0.70	&	1.33	&	0.91	&	10.5\%	&	0.09	&	0.70	&	1.84	&	1.59	&	7.8\%	&	0.13	&	1.32	&	0.54	&	0.14	&	1.5\%	&	0.68	\\
		&	\multicolumn{5}{c}{ALH-F4 $z_{max} = 2.4, N\approx4\cdot 10^4$}		&	\multicolumn{5}{c}{XMM$_{spec}$ $z_{max} = 3.6, N\approx1\cdot 10^3$} 	&	\multicolumn{5}{c}{zCOSMOS $z_{max} = 1.4, N\approx1\cdot 10^4$} 	\\
		0.05	&	0.81	&	2.40	&	1.14	&	0.3\%	&	3.10	&	0.62	&	1.44	&	1.23	&	0.1\%	&	7.99	&	1.05	&	2.71	&	0.75	&	0.3\%	&	3.62	\\
		0.1	&	0.82	&	2.41	&	1.14	&	1.0\%	&	1.00	&	0.72	&	1.26	&	1.01	&	0.3\%	&	2.86	&	1.10	&	2.58	&	0.72	&	1.1\%	&	0.91	\\
		0.2	&	0.73	&	2.53	&	1.19	&	2.8\%	&	0.36	&	0.76	&	1.22	&	0.94	&	1.4\%	&	0.74	&	0.92	&	3.06	&	0.79	&	3.7\%	&	0.27	\\
		0.3	&	0.63	&	2.80	&	1.28	&	12.4\%	&	0.08	&	1.51	&	0.78	&	0.27	&	2.2\%	&	0.45	&										\\
		
		\hline
	\end{tabular}
	\label{all_param}
\end{table*}
\noindent which provides an estimate of the observed dispersion of the fluctuations $\sigma_{obs} = |\sigma_{obs}|$ within $\Delta z$. In a standard $\Lambda$CDM model, the correlation function vanishes at 174 Mpc, and does not depend on the bias factor [18]. This means that bins of about 200 Mpc or more in size can be taken to be independent, and the signs of the fluctuations should alternate from bin to bin. However, in the presence of large-scale structure, the fluctuation sign will be preserved, and the number of neighboring bins displaying the same sign provides an estimate of the size of this structure.

\subsection{Predictions of the $\Lambda$CDM Model and the Galactic Bias Factor}
The standard $\Lambda$CDM model asserts that the correlation function for the density of visiblematter $\xi_{gal}$ (luminous baryonic matter) is related to the correlation function for the density of non-baryonic dark matter $\xi_{dm}$ by an additional the galaxy bias hypothesis
\begin{equation}
\label{xi-dm-gal}
\xi_{gal}(r, z, \pi) = b^2(z, \pi) \xi_{dm}(r, z),
\end{equation}

where $b^2(z, π)$ is the shift (bias) factor and $\pi = (L, m_{\star}, T, . . . )$.

In deep surveys, a substantial role is played by selection effects such as the Malmquist effect. Due to the decrease in the sensitivity of receivers with increasing distance to the objects studied, there is a systematic increase in the bias factor with redshift. The bias factor can be calculated using the formula [16]:

\begin{equation}
\label{bias}
b(m_\star ,\overline{z}) = b_0 (\overline{z}+1)^{b_1}+b_2
\end{equation}

where b0, b1, and b2 are parameters from [16, Table 4] that depend on the stellar mass of the galaxies. The sample-mean logarithm of the stellar mass of the galaxies in a bin was calculated using the formula

\begin{equation}
\label{lmass}
\log(m_\star) = \overline{\log (M/M_{\odot})} = \frac{1}{n}\sum_1^n { \log (M/M_{\odot}) }
\end{equation}

The relationship between the dispersions of the galaxy and dark-matter densities is given by the expression [16]

\begin{equation}
\label{gal_bias_dm}
\sigma_{gal}^2(m_\star, z) = b^2(m_\star, z)\sigma_{dm}^2(z),
\end{equation}

where $m_\star$ is the stellar mass of the galaxies and the dispersion of the dark-matter fluctuations is given by

\begin{equation}
 \label{Sigma_dm_2}
 \sigma_{dm}(z,\Delta z=0.2)=\frac{\sigma_a}{z^\beta+\sigma_b}
\end{equation}

where a, b, and $\beta$ are parameters from [16, Table 3] corresponding to the angular size of the catalog data for which the dispersion was calculated. The COSMOS and UVISTA catalogs have $\sigma_{a} = 0.069$, $\sigma_{b} = 0.234$, and $\sigma_{\beta} = 0.834$ for a bin size of $Δz = 0.2$. The following translation formula is used for other bin sizes:

\begin{equation}
 \label{Sigma_dm}
 \sigma_{dm}(z,\Delta z)=\sigma_{dm}(z,\Delta z=0.2)\sqrt{\frac{0.2}
 {\Delta z}} \;.
\end{equation}
\subsection{Reduced Dispersion}
Table 1 presents the results of calculating the reduced dispersion s using the formulas

\begin{equation}
s = \frac{\overline \sigma_{obs}}{\overline \sigma_{dm}},
  \label{rv}
\end{equation}

where

\begin{equation}
\overline \sigma_{obs}^2 = \frac {\Sigma(\sigma_{obs,i}-\overline \sigma_{obs,mean})^2}{n},
\label{line_sigma_obs}
\end{equation}

\begin{equation}
\overline \sigma_{dm}^2 = \frac {\Sigma(\sigma_{dm,i}-\overline \sigma_{dm,mean})^2}{n},
\label{line_sigma_dm}
\end{equation}

where $n$ is the number of bins and $\sigma_{obs}$, mean and $\sigma_{dm}$, mean correspond to the mean values over all the bins:

\begin{equation}
\overline \sigma_{obs,mean} = \frac {\Sigma(\sigma_{obs,i})}{n},
\label{mean_sigma_obs}
\end{equation}

\begin{equation}
\overline \sigma_{dm,mean} = \frac {\Sigma(\sigma_{dm,i})}{n}.
\label{mean_sigma_dm}
\end{equation}

\subsection{Correlation Coefficient}

To numerically compare the plots of the observed fluctuations in the number of galaxies $\delta_{obs}$, we calculated the linear correlation coefficient (or Pearson correlation coefficient) using the formula

\begin{equation}
  \rho_{XY} = \frac{cov_{XY}}{\sigma_X \sigma_Y} = \frac {
  \sum{(X-\overline{X})(Y-\overline{Y})}}
  {
  \sqrt{ \sum{(X-\overline{X})^2} \sum{(Y-\overline{Y})^2} }
  },
  \label{CC}
\end{equation}

where $\overline{X} = \frac {1}{n} \sum_{1}^n{X_i}$ and $\overline{Y} = \frac {1}{n} \sum_{1}^n{Y_i}$ are the mean values for the sample. The uncertainty $(\Delta_\rho )$ was calculated using a Fisher distribution:

\begin{equation}
\sigma_\rho = \sqrt {\frac{1-\rho^2}{n-2}},
  \label{sigma_CC}
\end{equation}


\section{DEEP FIELDS}

The method proposed in [8] can be used to estimate the scales of inhomogeneities in deep galaxy surveys using large redshift bins $(\Delta z = 0.05-0.3)$ that exceed the accuracy of the photometric redshift estimates $(\delta_z \sim 0.012(1 + z))$. Since such bins contain large numbers of galaxies $(N(\Delta z) \sim 10\,000)$, the Poisson noise $(\sim N^{-1/2})$ is small $(\sigma_P \sim 0.01)$, making it possible to detect fluctuations associated with large-scale inhomogeneities in the distribution of the galaxies.

We considered the following photometric surveys. The largest (more than 600\ 000 galaxes) and deepest ($z$ up to 5) surveys currently available are the COSMOS multi-band optical survey [19], the independent UltraVISTA infrared survey of the same field [20], and the ALH-F4 field [21], which is in this same area. We calculated the linear sizes and amplitudes of inhomogeneities in the radial distributions of the galaxies in the surveys, taking into account variations in the shapes of the spatial correlation functions for the galaxies and the appreciably nonspherical geometries of the samples. This comparison of the observed fluctuations with those predicted by the $\Lambda$CDM model indicates that the correlation structures have larger linear sizes and amplitudes than is expected in theoretical models for the evolution of non-baryonic dark matter, implying the need for large bias factor at high redshifts.

\begin{table}
	
	\caption{Comparison of the predicted deviations $\sigma_{gal}$ calculated using (12) and the observed fluctuations in the number of galaxies  $\sigma_{gal}$ in the UVISTA catalog for $\Delta z = 0.2$.}
	\begin{tabular}{llllllll}
		\hline \hline
		\multicolumn{8}{c}{UVISTA} \\
		\hline
		z	&	$lm_{\star}$	&	$b$	&	$\sigma_{dm}$	&	$\sigma_{gal}$	&	$\delta_{obs}$ & 5, 1.8 & 5, 1	\\
		0.1	&	7.68	&	-      	&	0.181	&	-	    &	$+0.034$	&	0.195	&	0.280		\\
		0.3	&	8.72	&	1.15	&	0.115	&	0.132	&	$+0.014$	&	0.114	&	0.211		\\
		0.5	&	9.16	&	1.25	&	0.087	&	0.109	&	$-0.189$	&	0.099	&	0.195		\\
		0.7	&	9.33	&	1.33	&	0.070	&	0.093	&	$-0.037$	&	0.092	&	0.186		\\
		0.9	&	9.49	&	1.43	&	0.060	&	0.086	&	$+0.215$	&	0.088	&	0.180		\\
		1.1	&	9.50	&	1.54	&	0.052	&	0.080	&	$+0.056$	&	0.085	&	0.175		\\
		1.3	&	9.60	&	1.74	&	0.047	&	0.082	&	$+0.050$	&	0.084	&	0.172		\\
		1.5	&	9.70	&	1.91	&	0.042	&	0.080	&	$+0.094$	&	0.082	&	0.169		\\
		1.7	&	9.85	&	2.13	&	0.039	&	0.083	&	$-0.098$	&	0.082	&	0.166		\\
		1.9	&	9.95	&	2.37	&	0.036	&	0.085	&	$-0.081$	&	0.081	&	0.163		\\
		2.1	&	9.96	&	2.66	&	0.033	&	0.088	&	$-0.207$	&	0.080	&	0.161		\\
		2.3	&	10.0	&	3.30	&	0.031	&	0.102	&	$-0.173$	&	0.080	&	0.159		\\
		2.5	&	10.1	&	3.71	&	0.029	&	0.108	&	$-0.046$	&	0.080	&	0.157		\\
		2.7	&	10.1	&	4.17	&	0.027	&	0.113	&	$+0.244$	&	0.079	&	0.155		\\
		2.9	&	10.1	&	4.68	&	0.026	&	0.122	&	$+0.034$	&	0.079	&	0.153		\\
		3.1	&	10.1	&	5.26	&	0.025	&	0.132	&	$+0.082$	&	0.079	&	0.152		\\
		3.3	&	10.0	&	5.89	&	0.024	&	0.141	&	$-0.024$	&	0.079	&	0.150		\\
		3.5	&	10.1	&	6.59	&	0.022	&	0.145	&	$-0.096$	&	0.078	&	0.148	    \\	
		
		\hline
	\end{tabular}
	\label{bias_UVISTA_F4}
\end{table}

We also analyzed the zCOSMOScatalog of spectroscopic redshifts [22], the XMM-COSMOS catalog of X-ray sources [23], the HDF-N catalog [24], and the ALH-F5 field [21]. Since these last two catalogs cover a different region of the celestial sphere, the pattern of the fluctuations in the number of galaxies they contain should differ from the pattern observed in the COSMOS/UVISTA field, as is shown by our results.

\subsection{COSMOS and UVISTA}
The COSMOS (Hubble Space Telescope Cosmic Evolution Survey) deep field is a pencil-beam ($1.3 \pm 1.3$ square degrees), multi-band survey [11, 25]. It includes 600\,000 galaxies and has a limiting magnitude $m_I < 26.6$. A sample of 385\,065 galaxies with $m_I < 25$ and photometric redshift uncertainties $\delta_z < 0.03$ for $z < 1.25$ and $\delta_z < 0.04-0.06$ for $z > 1.25$ is published in [19].

\begin{figure*}[h]
	\includegraphics[scale=0.66]{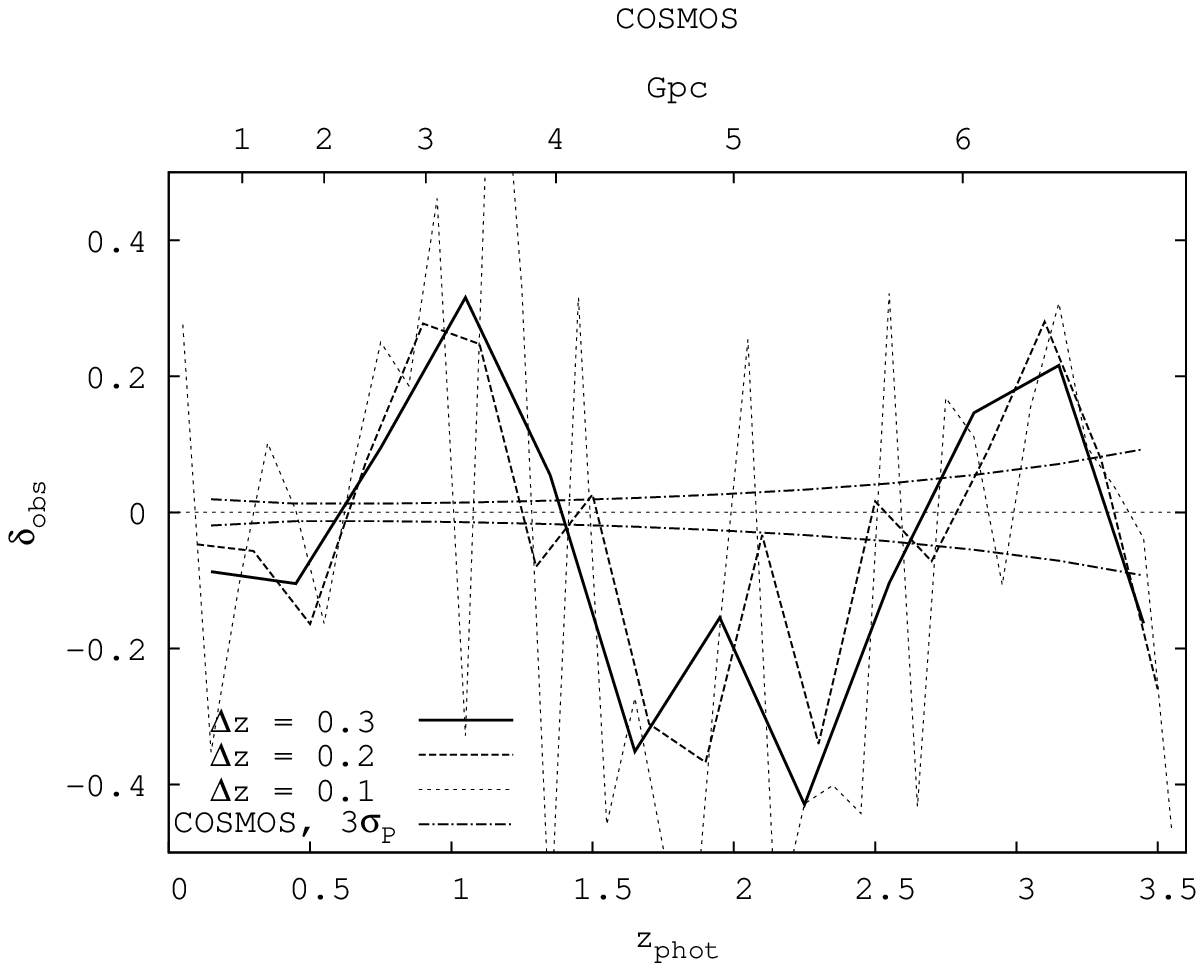}
	\hfill
	\includegraphics[scale=0.39]{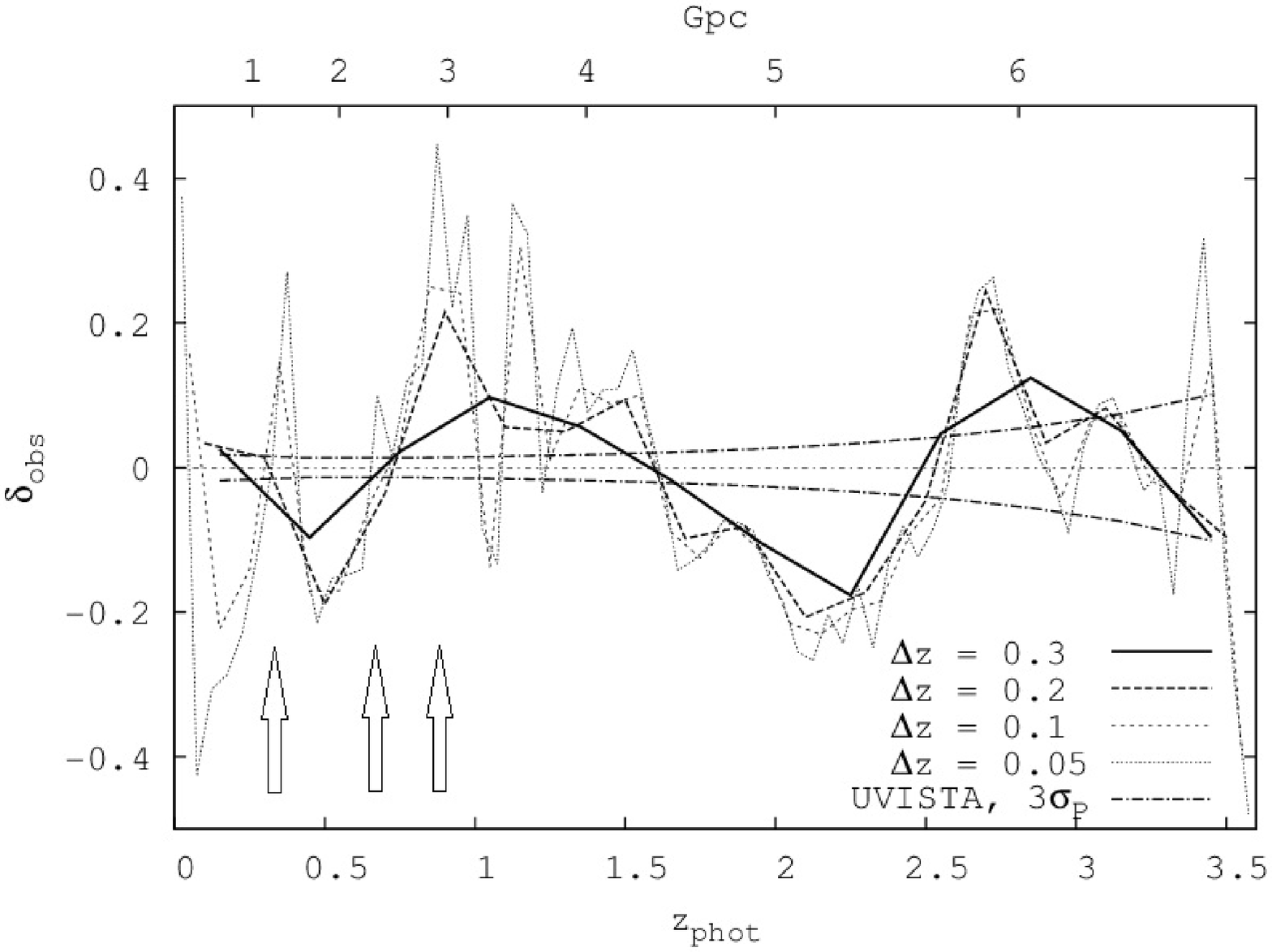}
	\caption{ Results of calculating the observed fluctuations in the number of galaxies $\delta_{obs}$ for the COSMOS (upper) and 	UltraVISTA (lower) catalogs for various redshift bins. The Poisson noise shown (dot-dashed curve) corresponds to 3$\sigma_P$	for $\Delta z = 0.3$. The arrows show redshifts for which clusters of galaxies have been detected in spectroscopic surveys.}
	\label{sigma_COSMOS_UVISTA}
\end{figure*}

\begin{table*}
	\centering
	\caption{
		Amplitudes $\delta_{str}$ and linear sizes $r$ (Mpc) of the estimated fluctuations for all catalogs with $\Delta$z $=0.3$.
	}
	\begin{tabularx}{\textwidth}{XXXXXXXXX}
		\hline  \hline
		Catalog & $z_s$	&	$z_f$	&	$r$	&	$N$	&	$3 \sigma_P$	&	$\delta_{str}$	&	5, 1	&	10, 1	\\
		\hline  															
		COSMOS	&	0.0	&	0.6	&	2206	&	71506	&	0.011	&	-- 0.10	&	0.16	&	0.23	\\[3pt]
		&	0.6	&	1.4	&	1968	&	135429	&	0.009	&	+ 0.14	&	0.11	&	0.15	\\[3pt]
		&	1.4	&	2.7	&	1877	&	45489	&	0.013	&	-- 0.17	&	0.08	&	0.12	\\[3pt]
		&	2.7	&	3.3	&	578	    &	5583	&	0.044	&	+ 0.18	&	0.11	&	0.16	\\[3pt]
		\hline  													
		UVISTA	&	0.2	&	0.7	&	1689	&	72579	&	0.011	&	-- 0.10	&	0.14	&	0.20	\\[3pt]
		&	0.7	&	1.6	&	2038	&	119870	&	0.009	&	+ 0.08	&	0.10	&	0.14	\\[3pt]
		&	1.6	&	2.5	&	1284	&	32525	&	0.016	&	-- 0.11	&	0.09	&	0.13	\\[3pt]
		&	2.5	&	3.3	&	801	    &	8351	&	0.035	&	+ 0.11	&	0.10	&	0.14	\\[3pt]
		\hline  													
		zCOSMOS	&	0.13	&	0.39	&	977	&	2581	&	0.064	&	+ 0.17	&	0.20	&	0.29	\\[3pt]
		&	0.39	&	0.67	&	901	&	2403	&	0.051	&	-- 0.31	&	0.17	&	0.25	\\[3pt]
		&	0.67	&	1.01	&	910	&	3219	&	0.058	&	+ 0.21	&	0.15	&	0.21	\\[3pt]
		&	1.01	&	1.27	&	584	&	365		&	0.139	&	-- 0.22	&	0.16	&	0.23	\\[3pt]
		\hline  													
		XMM$_{phot}$	&	0.22	&	0.56	&	1188	&	204	&	0.206	&	-- 0.04	&	0.17	&	0.24	\\[3pt]
		&	0.56	&	1.06	&	1366	&	457	&	0.146	&	+ 0.08	&	0.13	&	0.18	\\[3pt]
		&	1.06	&	1.56	&	1026	&	383	&	0.151	&	-- 0.03	&	0.12	&	0.17	\\[3pt]
		&	1.56	&	2.04	&	764		&	282	&	0.181	&	+ 0.03	&	0.12	&	0.17	\\[3pt]
		&	2.04	&	2.46	&	544		&	146	&	0.246	&	-- 0.02	&	0.13	&	0.18	\\[3pt]
		&	2.46	&	2.97	&	546		&	110	&	0.306	&	+ 0.15	&	0.12	&	0.16	\\[3pt]
		&	2.97	&	3.38	&	370		&	34	&	0.507	&	-- 0.03	&	0.12	&	0.17	\\[3pt]
		\hline 														
		ALH-F4	&	0.20	&	0.60	&	1390	&	10202	&	0.031	&	+ 0.08	&	0.17	&	0.24	\\[3pt]
		&	0.60	&	0.83	&	663		&	6212	&	0.036	&	-- 0.09	&	0.19	&	0.27	\\[3pt]
		&	0.83	&	1.16	&	807		&	10692	&	0.032	&	+ 0.20	&	0.16	&	0.23	\\[3pt]
		&	1.16	&	1.64	&	937		&	6325	&	0.035	&	-- 0.15	&	0.14	&	0.19	\\[3pt]
		&	1.64	&	2.15	&	776		&	2167	&	0.068	&	+ 0.12	&	0.13	&	0.19	\\[3pt]
		\hline														
		ALH-F5	&	0.53	&	0.97	&	1244	&	3305	&	0.049	&	-- 0.11	&	0.15	&	0.21	\\[3pt]
		&	0.97	&	1.58	&	1278	&	3676	&	0.053	&	+ 0.13	&	0.13	&	0.18	\\[3pt]
		&	1.58	&	2.17	&	908		&	581		&	0.116	&	-- 0.14	&	0.13	&	0.18	\\[3pt]
		\hline														
		HDF-N	&	0.21	&	0.55	&	1194	&	280	&	0.203	&	+ 0.28	&	0.19	&	0.27	\\[3pt]
		&	0.55	&	0.92	&	1055	&	163	&	0.180	&	-- 0.41	&	0.17	&	0.23	\\[3pt]
		&	0.92	&	1.54	&	1332	&	515	&	0.147	&	+ 0.23	&	0.13	&	0.18	\\[3pt]
		&	1.54	&	1.81	&	456		&	149	&	0.243	&	-- 0.03	&	0.10	&	0.14	\\[3pt]
		&	2.26	&	2.88	&	704		&	146	&	0.198	&	-- 0.37	&	0.14	&	0.19	\\[3pt]
		&	2.88	&	3.36	&	442		&	178	&	0.258	&	+ 0.32	&	0.16	&	0.22	\\
		\hline
	\end{tabularx}
	\label{tab_sigma_obs}
\end{table*}

We applied a stronger constraint on the quality of the photometric redshifts: the presence of a single peak in the probability distribution for the measured photometric redshift. The total number of galaxies in the first sample is 258\,491 at $z < 3.6$, while the number in the second sample is 239\,750 at $z < 2.4$. The samples with different limiting redshifts zmax demonstrate how the fluctuation method behaves as the depth of the sample changes, and make it possible to more correctly compare the results of the fluctuation method for catalogs with different limiting magnitudes (and thus different zmax values).

\begin{table*}
	\centering
	\caption{Correlation coefficients $\rho$, their uncertainties $\sigma_\rho$, and the significances $P=1-\alpha$ according to a Student’s distribution for a pairwise comparison of the catalogs ($\Delta z=0.2$ if not noted otherwise)}
	\begin{tabularx}{\textwidth}{XXXXXXXXX}
		\hline	
		\multicolumn{3}{c}{pair catalogs}	&	$z_{max}$	&	$\rho$	&	$\sigma_\rho$  &   $P$	\\
		\hline											
		\multicolumn{3}{c}{COSMOS \& ALH-F4}	      	&	1.7	&	+ 0.53	&	0.38   &   0.9	\\[3pt]
		\multicolumn{3}{c}{COSMOS \& UVISTA}			    &	3.6	&	+ 0.69	&	0.24   &   0.99	\\[3pt]
		\multicolumn{3}{c}{UVISTA \& ALH-F4}			    &	1.7	&	+ 0.59	&	0.36   &   0.9	\\[3pt]
		\multicolumn{3}{c}{UVISTA \& COSMOS}			    &	2.4	&	+ 0.70	&	0.32	&   0.95\\[3pt]
		\multicolumn{3}{c}{COSMOS \& zCOSMOS $\Delta z=0.05$}	&	1.4	&	+ 0.52	&	0.21   &   0.98	\\[3pt]
		\multicolumn{3}{c}{COSMOS \& zCOSMOS $\Delta z=0.1$}	&	1.4	&	+ 0.74	&	0.26   &   0.98    \\[3pt]
		\multicolumn{3}{c}{COSMOS \& zCOSMOS $\Delta z=0.2$}	&	1.4	&	+ 0.79	&	0.43	&   0.9     \\[3pt]
		\multicolumn{3}{c}{XMM$_{phot}$ \& C+U}			&	3.6	&	+ 0.55	&	0.25	&   0.975     \\[3pt]
		\multicolumn{3}{c}{XMM$_{spec}$ \& C+U}				&	3.6	&	+ 0.51	&	0.27	&   0.95      \\[3pt]
		\multicolumn{3}{c}{XMM$_{phot}$ \& C+U+F4}			&	1.7	&	+ 0.74	&	0.30	&   0.97     \\[3pt]
		\multicolumn{3}{c}{XMM$_{spec}$ \& C+U+F4}			&	1.7	&	+ 0.82	&	0.20	&   0.998    \\[3pt]
		\multicolumn{3}{c}{XMM$_{phot}$ \& COSMOS}			&	3.6	&	+ 0.51	&	0.26	&   0.96    \\[3pt]
		\multicolumn{3}{c}{XMM$_{spec}$ \& COSMOS}			&	3.6	&	+ 0.60	&	0.27	&   0.975   \\[3pt]
		\multicolumn{3}{c}{XMM$_{phot}$ \& UVISTA}			&	3.6	&	+ 0.55	&	0.25	&   0.975   \\[3pt]
		\multicolumn{3}{c}{XMM$_{spec}$ \& UVISTA}			&	3.6	&	+ 0.37	&	0.28   &   0.9   \\[3pt]
		\multicolumn{3}{c}{XMM$_{phot}$ \& F4}				&	1.7	&	+ 0.82	&	0.29	&   0.975  \\[3pt]
		\multicolumn{3}{c}{XMM$_{spec}$ \& F4}				&	1.7	&	+ 0.65	&	0.38   &   0.9  \\[3pt]
		\multicolumn{3}{c}{HDF-N \& COSMOS}					&	3.6	&	-- 0.20	&	0.31 & --  \\[3pt]
		\multicolumn{3}{c}{CCD-1 \& HDF-N $\Delta z=0.1$}		&	1.7	&	+ 0.61	&	0.21    &   0.99	\\[3pt]
		\multicolumn{3}{c}{CCD-1 \& HDF-N $\Delta z=0.2$}		&	1.7	&	+ 0.61	&	0.32    &   0.95 	\\[3pt]
		\multicolumn{3}{c}{CCD-1 \& HDF-N $\Delta z=0.3$}		&	1.7	&	+ 0.85	&	0.31    &   0.96	\\
		\hline
	\end{tabularx}
	
	\label{CC_tab}
\end{table*}

\begin{figure}
	\includegraphics[scale=0.66]{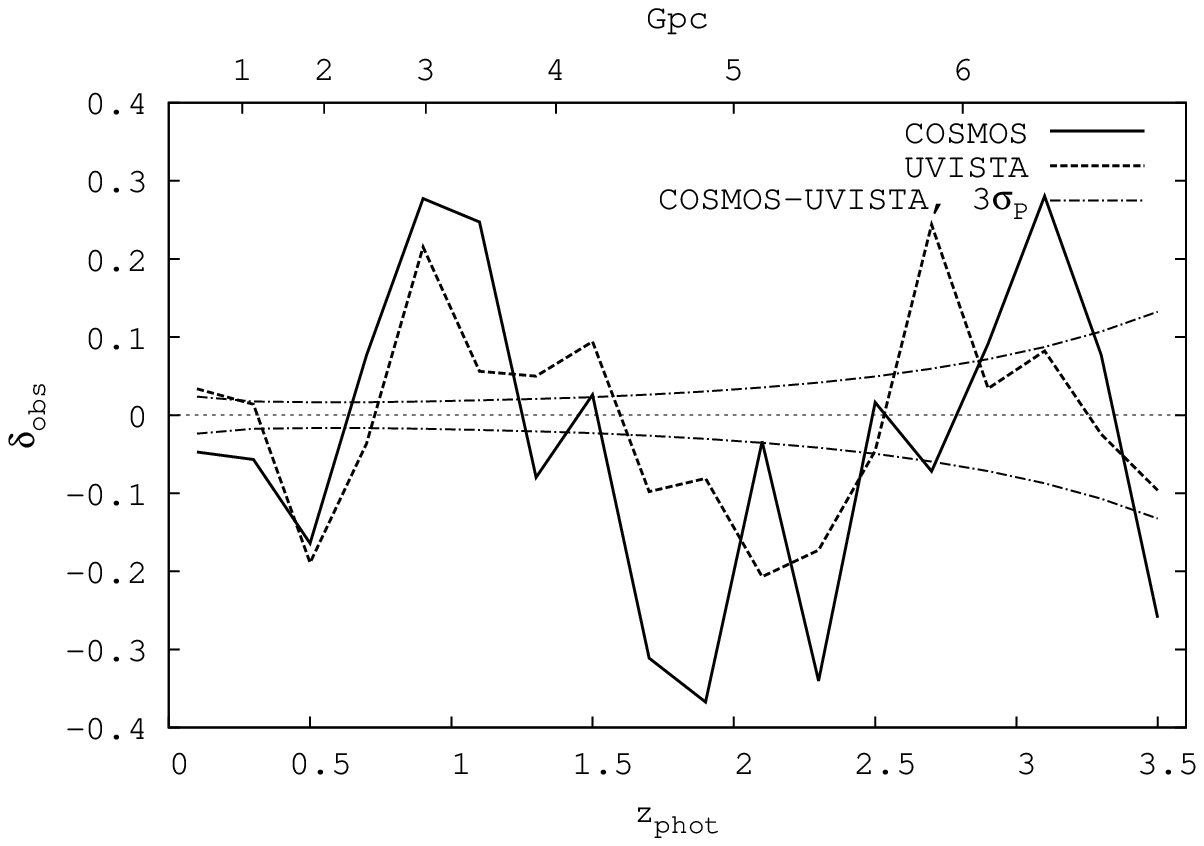}
	\vfill
	\includegraphics[scale=0.66]{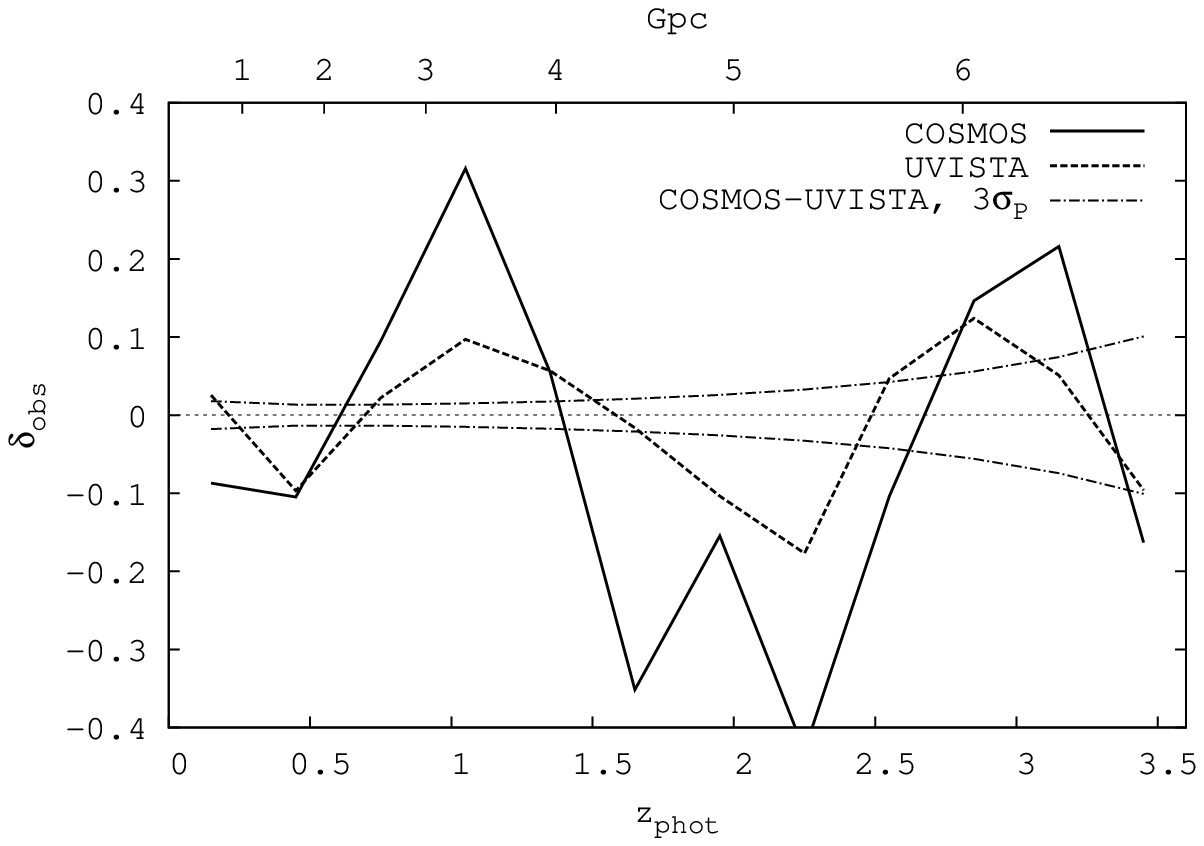}
	\caption{ Pairwise comparison of the observed deviations δobs for $\delta_{obs}$ (upper) and $\Delta z =0.2$ (lower) for the COSMOS (solid curve) and UVISTA (dotted curve) catalogs. An obvious correlation of the deviations δobs in the interval $z = 0.2−1.5$ can be seen. The sign of the fluctuations is negative at $z \sim 0.5$ (a minimum) and positive at $z \sim 1$ (a maximum).}
	\label{C+U}
\end{figure}

The UltraVISTA survey (Ultra Deep Survey with the VISTA telescope) [20], based on GALEX, Subaru, CFHT, VISTA (European Southern Observatory) and Spitzer data, contains photometric redshifts for 262\,615 sources with magnitudes $K_s < 24.35$. The accuracy of the photometric redshifts is $\sim 0.013(1 + z)$ (the corresponding accuracy for COSMOS is $\sim 0.012(1 + z)$. The parameters of the angular extent of the catalog were specially made to coincide with those for COSMOS, to facilitate comparisons of the two. An additional criterion for inclusion in the survey was the probability that the photometric redshifts were correct: objets for which $"peakprob" > 0.9$ were included. As for the COSMOS catalog, two samples were composed: 250\ 769 objects with $z < 3.6$ and 247\ 594 objects with $z < 2.4$.

The uniqueness of the COSMOS and UVISTA deep surveys is that each of them contains more than 200\,000 uniformly selected galaxies with measured fluxes in 30 filters, making it possible to determine their photometric redshifts with accuracies $\delta_z < 0.1$. This is sufficient for studies of inhomogeneities in the galaxy distributions on scales $\Delta z > 0.1$; i.e., $\Delta R > 300$ Mpc/h. The results in Fig. 2 show that decreasing the bin size leads to a noisier image for the COSMOS catalog and to a more realistic fluctuation map for the UVISTA catalog (with increased resolution). 

Note that the peak at $z = 0.73$ for $\Delta z = 0.05$ corresponds to a galaxy cluster detected using spectroscopic observations [26], and the observed three peaks at $z \sim 0.35$, $z \sim 0.7$, and $z \sim 0.85$ correspond to the results of [22]. Figure 3 shows the correlation between the signs of the fluctuations for these two independent catalogs.

The least-squares parameters ($\alpha, \beta, z_0$) and the sum of the squared deviations ($\Sigma$) for all samples and for all bins are presented in Table 2. The combined COSMOS \& UVISTA catalog is denoted C\&U. The quantity $\delta = \epsilon /\Sigma$, where $\epsilon = 0.01$ is the accuracy of the fitted parameters. The parameter A in (4) is normalized to the total number of galaxies in the sample (N).

Table 3 gives a comparison of the observed and	theoretical fluctuations in the numbers of galaxies for the UVISTA catalog for $\Delta z = 0.2$. Here, $z$ is the mean redshift for a bin, $\log(m_\star)$ the logarithm of the stellar mass, $b$ the bias factor, and $\sigma_{dm}$ the maximum amplitude of the dark-matter fluctuations. The predicted fluctuations in the number density of galaxies $\sigma_{gal}$ are less than half $\sigma_{obs}$ at redshifts $z \sim 0.9$ and $z \sim 2.1$. Deficits and excesses of galaxies are present at these redshifts, as will be shown below. The last two columns of this table give the results of numerically integrating (6) with the correlation function (7) for the parameters $(r0; \gamma) = (5;1.8)$ and $(5; 1)$. The 20\% relative amplitude of the fluctuations in the COSMOS/UVISTA field can be explained if the parameters of the spatial correlation function are $(\gamma; r0) = (1;5-10)$. Note that just such parameters for the power-law correlation function were found in the 2dF and SDSS wide-angle modern redshift surveys [27]. Table 4 shows the amplitudes and linear sizes of the fluctuation estimates for all the catalogs for $\Delta z = 0.3$. The linear size $r$ (in Mpc) was taken to be the difference between the two metric distances $r(z_s)$ and $r(z_f)$, calculated using (1). The number of visible objects in a given interval is $N$. The Poisson noise has a level of $3\sigma_P$ . The amplitude $\delta_{str}$ was calculated using (8). The last two columns of this table show the results of numerically integrating (6) with the correlation function (7) for $(r0; \gamma) = (5;1); (10; 1)$.

\begin{figure}
	\includegraphics[scale=0.6]{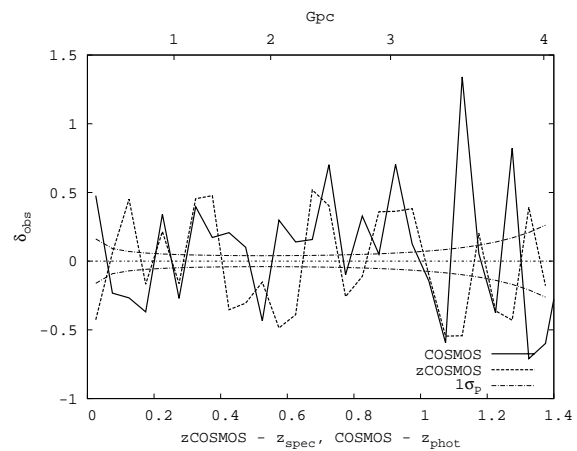}
	\caption{ Comparison of the observed deviations $\delta_{obs}$ of COSMOS (solid) and 10k zCOSMOS (dotted) catalogs for $\Delta z=0.05$. Poisson noise for the COSMOS is 1$\sigma_P$ at $\Delta z = 0.05$ marked by dot-dash lines.}
	\label{sigma_zCOSMOS_COSMOS}
\end{figure}

The resulting correlation coefficients are presented in Table 5. A working interval, shortened at the ends, was distinguished, in which the correlation coefficient
was calculated. The point is that the fit may be inaccurate at the limiting z values due to the small numbers of objects in these bins. At the beginning of the interval, this is due to the small volume of the subsample, and at the end of the interval, to the Malmquist effect. Thus, one to two points on the left and four to seven points on the right -- i.e., roughly 25\% of the total number of objects in the sample-were rejected when calculating the correlation coefficient.

\subsection{zCOSMOS, XMM-COSMOS and ALHAMBRA-F4}

The ALHAMBRA (Advance Large Homogeneous Area Medium Band Redshift Astronomical) survey [21] encompasses eight different regions of the sky, including sections of the SDSS, DEEP2, ELAIS, GOODS-N, COSMOS, and GROTH fields. It uses a new photometric system with 20 adjacent filters and a $\sim 300$\AA{} transmission bandwidth covering the optical range, together with deep JHK images. These observations were carried out on the 3.5 m Calar Alto telescope (Spain). The catalog contains $\sim$ 438\,000 galaxies with apparent magnitudes $m_I < 24.5$. 

The 30'$\times$30' ALHAMBRA-Field 4 field (four essentially adjacent 15'$\times$15' frames) containing 37\,854 objects was taken for comparison with the COSMOS and UVISTA surveys. The $"stellarflag"$ parameter, which is equal to unity if an object is a star and is less than unity if it a galaxy, was adopted as a selection parameter for the sample. We selected 36\,627 sources with photometric redshifts $z < 2.4$ and $stellarflag < 0.9$. The same criteria
were applied to the ALHAMBRA-Field 5 field, which is not far from the HDF-N field [24]. We used only two of four frames for ALH-F5 (denoted CCD-1 and CCD-2), containing 10\,064 and 10\,655 objects at $z < 2.4$, respectively.

The 10k-zCOSMOS catalog of spectroscopic redshifts [22], which contains $\sim 105$ sources and has $15 < m_{I_{AB}} < 22.5$, falls in the COSMOS/UVISTA field. Observations were carried out using the VIMOS spectrograph over 600 hrs on the 8-m Very

\begin{figure}
\includegraphics[scale=0.66]{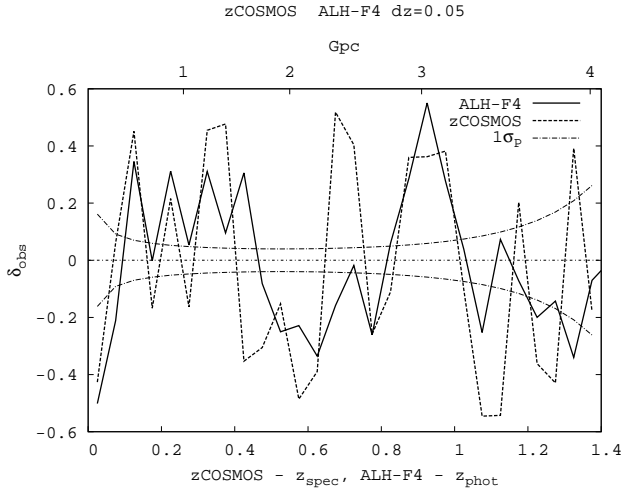}
\caption{ 	Comparison of the observed deviations $\delta_{obs}$ for COSMOS (solid curve) and 10k–zCOSMOS (dotted curve) for	$\Delta z=0.05$. The Poisson noise shown (dot-dashed curve) corresponds to 1$\sigma_P$ for the COSMOS catalog for $\Delta z = 0.05$.}
\label{sigma_zCOSMOS_ALH-F4}
\end{figure}

\begin{figure}
\includegraphics[scale=0.66]{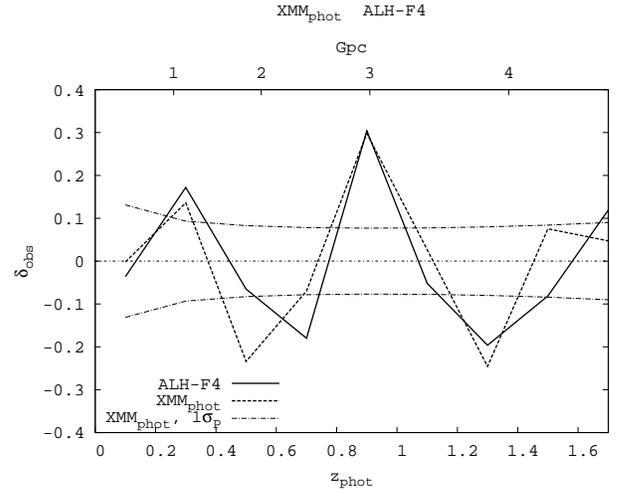}
\caption{Comparison of $\delta_{obs}$ for the XMM photometric catalog (dotted curve) and the ALH-F4 sample (solid curve) for $\Delta z = 0.2$.	The Poisson noise shown (dot-dashed curve) corresponds to 1$\sigma_P$ for the catalog for $\Delta z = 0.2$}
\label{sigma_XMM_F4}
\end{figure}
\begin{figure}
	\includegraphics[scale=0.66]{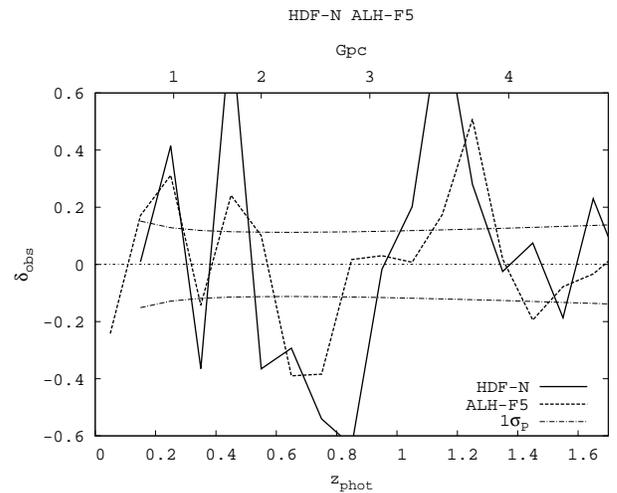}
	\caption{Comparison of $\delta_{obs}$ for the HDF-N catalog (solid curve) and the ALH-F5/CCD-1 sample (dotted curve) for $\Delta z = 0.1$. The Poisson noise shown (dot-dashed curve) corresponds to 1σP for the HDF-N catalog. }
	\label{HDF-N_F5_sigma}
\end{figure}
\begin{figure*}
\includegraphics[scale=0.66]{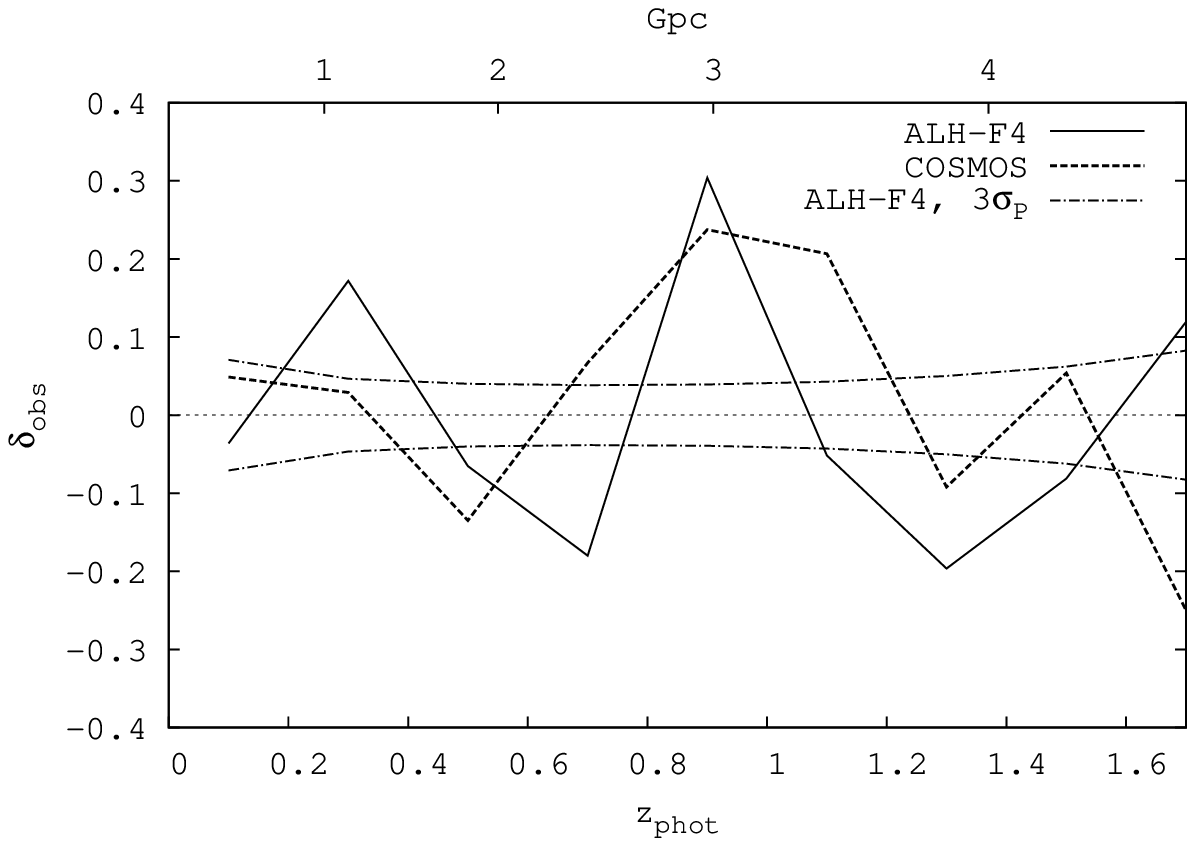}
\hfill
\includegraphics[scale=0.66]{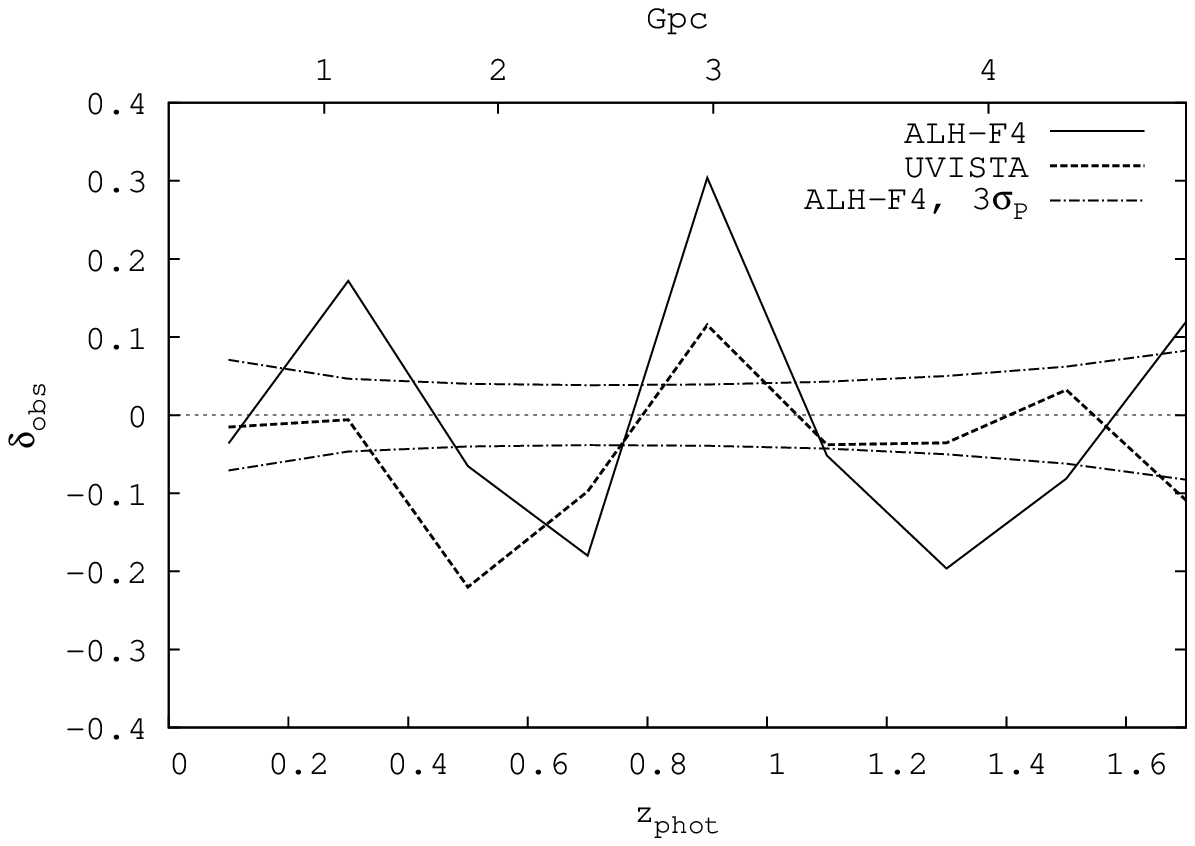}
\caption{ Pairwise comparison of the observed deviations $\delta_{obs}$ for the COSMOS (dotted curve) and ALH-F4 (solid curve) catalogs (upper) and the UVISTA (dotted curve) and ALH-F4 (solid curve) catalogs (lower), for $\Delta z =0.2$. A correlation between the δobs deviations in the interval $z = 0.2−1.5$ can be seen. The fluctuations are negative at $z \sim 0.6$ (a minimum) and positive at$z \sim 0.9$ (a maximum).}
\label{C_U_F4}
\end{figure*}

Large Telescope of the European Southern Observatory. This survey covers 1.7 square degrees on the celestial sphere, and coincides with the COSMOS field. The sample contains 9\,167 galaxies with known redshifts in the interval $0.1 < z < 1.4$. 

The XMM-Newton Wide-Field Survey in the COSMOS field (or simply XMM-COSMOS) catalog [23] contains $\sim$1\,800 X-ray source with fluxes greater than $\sim 5 \times 10^{-16}$, $\sim 3 \times 10^{-15}$, $\sim 7 \times 10^{-15}$ erg cm$^-2$ s$^-1$ at $0.5-2$, $2-10$, and $5-10$ keV, respectively. This catalog uses data taken from the COSMOS survey. Two samples were produced, the first containing 1\,666 high-quality photometric redshifts and the second containing 844 spectroscopic redshifts, both out to $z_{max} = 3.6$. Figures 4 and 5 show the correlations of the fluctuation signs for zCOSMOS with COSMOS and ALH-F4. Figure 6 depicts the correlation of the fluctuations between the XMM photometric catalog and the ALH-F4 sample. Figure 7 compares the fluctuation patterns for the ALH-F4 sample with the COSMOS and UVISTA catalogs.
\section{Northern Hubble Deep Field}
The HDF-N catalog and ALH-F5 sample were taken for comparison with the COSMOS field. 

The ALHAMBRA-F5 survey contains four images separated by about one degree. For definiteness, we took the first image (CCD-1), which contains 10\,510 objects. We made a subsample of these with $z_{max} = 1.8$ containing 9\,827 objects. We made a subsample of objects with high-quality photometric redshifts (with the probability of determining the first peak $>70$\%) with redshifts to $z_{max} = 3.6$ from the
HDF-N catalog; this subsample contains 1\,761 objects.

The corresponding fitted coefficients describing the radial distribution of galaxies in the HDF-N and ALH-F5 catalogs are presented in Table 2. The correlation of the fluctuations for HDF-N and ALHF5 shown in Fig. 8 demonstrates similar behavior for the inhomogeneities in methodologically different independent surveys of the same deep field. An anti-correlation between the radial distributions of the galaxies in the COSMOS and HDF-N surveys is shown in Fig. 9. This demonstrates an
absence of universal selection effects that act in the same way in all surveys. This also implies that the inhomogeneities are already independent on angular

\section{Conclusions}
Our analysis of the radial distribution of galaxies in the COSMOS/UVISTA field shows that the real observed fluctuations in the spatial distribution of the galaxies appreciably exceed the predictions of the $\Lambda$CDM model for the evolution of non-baryonic dark matter, in both their amplitude and linear size. This means that the $\Lambda$CDM model requires the introduction of a large bias factor ($b\sim10$ relative to the non-baryonic dark matter) at redshifts  $z\sim1$. It is also necessary to explain the large scale of the positive correlation corresponding to the linear size of the detected structures. This follows from Table 3, which shows that the fluctuations in the number of galaxies preserve their sign over several adjacent bins, while neighboring bins should have opposite signs in the $\Lambda$CDM model. Thus, in addition to the difficulties of the $\Lambda$CDM model on small scales (galaxies and halos with sizes of $10-100$ kpc [28, 29]), there also exist problems on very large scales, associated with the presence of large-scale inhomogeneities in the spatial distribution of galaxies with sizes of the order of 1\,500 Mpc and amplitudes exceeding 20\%.

\begin{figure}
	\centering
	\includegraphics[scale=0.66]{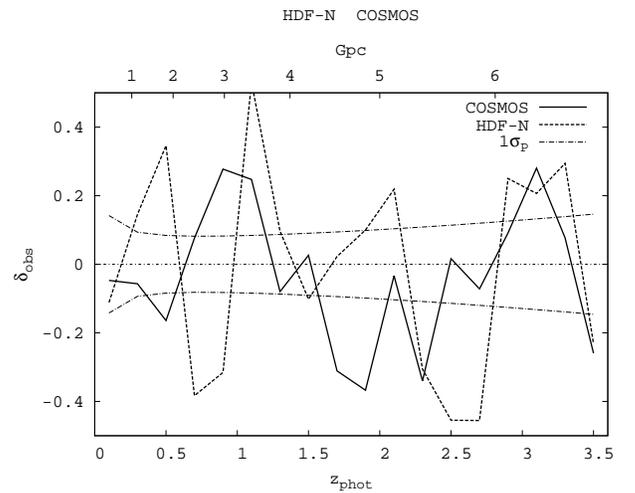}
	\caption{Comparison of $\delta_{obs}$ for the COSMOS (solid curve) and HDF-N (dotted curve) catalogs. The Poisson noise shown	(dot-dashed curve) corresponds to $\sigma_{P}$ for the HDF-N catalog.}
	\label{HDF-N_COSMOS}
\end{figure}

Our analysis of the COSMOS and UVISTA survey data and comparisons of these data with the data from the ALHAMBRA, zCOSMOS, and XMMCOSMOS surveys leads to the following conclusions. 
\begin{itemize}
  \item The detected inhomogeneities in the radial distribution of galaxies in the COSMOS, UVISTA, ALHAMBRA, and XMMphot-COSMOS photometric catalogs is confirmed by data from the zCOSMOS and XMMspec-COSMOS spectroscopic catalogs, and these data are mutually consistent.
  \item The amplitudes and linear sizes of the fluctuations in the independent COSMOS (optical) and UltraVISTA (near infrared) catalogs, and also in the ALHAMBRA/Field 4, XMMNewton and zCOSMOS catalogs, are mutually consistent. The corresponding correlation coefficient is positive and equal to $\rho > 0.5$.
  \item The amplitudes and sizes of the fluctuations are stable for different fits and various limiting redshifts $z_{max}$. When the bin size is decreased, the fluctuation amplitude grows, and the individual density peaks coincide with galaxy clusters detected earlier.
\end{itemize}
The amplitudes and sizes we have found agree with the amplitudes and sizes of inhomogeneities found for the COSMOS field in other studies using the 10k-zCOSMOS spectral survey [22, 30], ALHAMBRA photometric survey [21], and X-ray observations [31].

Appreciable fluctuations in the number density of galaxies in slices of the COSMOS survey at various redshifts were found in [32], where is it emphasized that these structures really exist. Thirty-six candidate structures were found at redshifts $1.5 < z < 3.1$, having masses of $10^{15}M_{\odot}$. The sizes of the observed radial structures appreciably exceed the transverse cross sections of the pencil-beam surveys. The detected individual galaxy clusters fall near peaks of the fluctuations found using small redshift-bin sizes [25].

In particular, the peak at $z = 0.73$ corresponds to a galaxy cluster that was detected in [26] using spectroscopic observations, and the three peaks at $z\sim0.35$, $z\sim0.7$, and $z\sim0.85$ coincide with clusters detected in [22], as well as in our own study. The paper [21] describes the ALHAMBRA catalog, which includes the region of the COSMOS survey. This catalog also displays non-uniformity in the radial distribution of galaxies, which is correlated with the non-uniformity observed for the COSMOS survey. X-ray sources from the COSMOS catalog are considered in [31]. Peaks in the radial distribution of these X-ray sources agree with regions where there are excess galaxies in the optical and IR, indirectly supporting the presence of large-scale structures. The detection of fluctuations in the number of galaxies, manifest in the same way in independent observations and obtained using independent datareduction methods, substantially reduces the possibility that these are associated with unknown systematic errors. This suggests with a high degree of certainty that the fluctuations observed in the COSMOS/UVISTA field are related to the cosmic variance, and thus imply positive correlations in the spatial distribution of galaxies in deep surveys.

\subsection*{Acknowledgments}
This work was supported by a grant from St. Petersburg State University (No. 6.38.18.2014). We are grateful for the opportunity to use the COSMOS\footnote{http://cosmos.astro.caltech.edu/}, zCOSMOS\footnote{http://archive.eso.org/cms/eso-data/data-packages/zcosmos-data-release-dr1.html}, XMM-Newton\footnote{http://xmmssc-www.star.le.ac.uk/Catalogue/ xcat\_public\_3XMM-DR4.html}, UVISTA\footnote{http://www.strw.leidenuniv.nl/galaxyevolution/ULTRAVISTA/\\Ultravista/K-selected.html}, ALHAMBRA\footnote{https://cloud.iaa.csic.es/alhambra/} and HDF-N\footnote{http://www.stsci.edu/ftp/science/hdf/hdf.html}.


\label{lastpage}

\end{document}